\documentclass[twocolumn,showpacs,preprintnumbers,amsmath,amssymb]{revtex4-2}
\usepackage{bm}

\usepackage{hyperref}
\usepackage{graphicx}
\usepackage{mwe}
\usepackage{subfigure}
\usepackage{dcolumn}
\usepackage{xcolor}

\usepackage{float}
\usepackage[capitalize]{cleveref}
\hypersetup{
    colorlinks=true,
    linkcolor=blue,
    citecolor=blue, allcolors=blue
}
\usepackage{figsize}
\usepackage{mathtools}

\begin{document}
\title{Chiral Tunneling through Double Barrier Structure in Twisted Graphene Bilayer}

\vspace{5mm}

\author{ A. Bahlaoui$^{1}$ and Y. Zahidi$^{1}$ }

\affiliation{$^{1}$\em Multidisciplinary Research and Innovation Laboratory, Polydisciplinary Faculty, Sultan Moulay Selimane University, 25000 Khouribga, Morocco}

\begin{abstract}

The paper discusses the chiral tunnelling of charge carriers through double barrier structure in twisted graphene bilayer. The theoretical analysis investigates the transmission probability for various system parameters under both symmetric and asymmetric barrier conditions. The results reveal that the transmission probability of quasiparticles in the $K$ cone is mirror symmetric to that of $K_{\theta}$ cone about $\varphi =0$. Furthermore, the study shows that the transmission changes gradually from perfect transmission to perfect reflection in the normal direction by increasing the incident energy and the barrier height, which is different from the case of monolayer and AB-stacked bilayer graphene. It is also found that the double barrier structure remains, only in certain cases, perfectly transparent for normal or near-normal incidence. The chiral nature of the quasiparticles in graphene causes the tunneling to be highly dependent on the direction and also on the double barrier structure. Interestingly, this characteristic provides additional parameter that allows us to tune the electronic properties of the twisted graphene bilayer. Additionally, we found that the transmission exhibits some sharp resonance peaks, the number and amplitude of which depend on the system parameters. Our results provide a better understanding of the chiral tunnelling in twisted graphene bilayer through double barrier structures and can help in designing efficient electronic devices.


\end{abstract}
\maketitle
\section{INTRODUCTION}

Graphene \cite{ref2} is a two-dimensional material composed of carbon atoms arranged in a hexagonal lattice. Due to its fascinating properties, it has attracted immense interest from researchers \cite{zhang2005experimental, peres2006electronic,katsnelson2006chiral, bai2007klein}. Charge carriers in this material have an approximately linear electron spectrum near the Fermi energy at two inequivalent points of the Brillouin zone (BZ), resulting in massless, chiral “relativistic” fermions governed by a two-dimensional Dirac equation \cite{beenakker2008colloquium, neto2009electronic}. This leads to a number of unusual electronic properties such as high electron mobility and quantum Hall effect \cite{novoselov2005two}. The presence of Dirac-type quasiparticles in graphene is also expected to lead to perfect transmission through potential barriers, a phenomenon known as Klein tunneling \cite{geim2007rise}. To gain insight into the transmission and transport properties of graphene-based structures, it is important to have a means of creating potential barriers on graphene. Various theoretical approaches, such as the use of electrostatic or magnetic fields \cite{pereira2007graphene, pereira2010klein, chen2009design}, breaking-symmetry substrates \cite{masir2009magnetic, de2007magnetic}, and mechanical deformation \cite{cayssol2009contact, gattenlohner2010dirac}, can be used to achieve this goal.\\

The chiral nature of graphene is a fascinating property that has the potential to revolutionize nanoelectronics and related fields. In addition to the Klein tunneling phenomenon, the conservation of the carriers' pseudospin in graphene could have significant implications for electronic and optical device applications. For example, the chiral nature of graphene could lead to the development of novel optoelectronic devices, such as circularly polarized light sources and detectors. These devices could offer improved performance and new functionality compared to their traditional counterparts. Furthermore, the ability to manipulate the chirality of graphene could allow for the design of advanced electronic devices with unique properties. To investigate such phenomena, researchers have investigated different graphene-based microstructures, including n-p junctions \cite{cheianov2006selective}, single and double barriers \cite{katsnelson2006chiral, pereira2007graphene}, and superlattices \cite{bai2007klein}. The study of these microstructures is essential to further understanding the fundamental physics of graphene and its potential applications. Overall, graphene's unique properties and potential applications make it a highly promising material for use in electronic, photonic, and sensor devices, and it is likely to continue to be the focus of intense research in the near future.\\

Twisted graphene bilayer (TGB) is a type of graphene consisting of two layers rotated with respect to each other at a specific "magic" angle. The angle of rotation determines the size of the Moiré pattern, which is a superlattice structure that arises from the overlapping of the two graphene lattices. This Moiré pattern can significantly affect the electronic properties of TGB, giving rise to flat bands and correlated insulating phases at certain angles \cite{cao2018unconventional,cao2018correlated}. One of the most fascinating electronic properties of TGB is the presence of chiral tunneling, which arises from the conservation of pseudospin in chiral particles and leads to the suppression of backscattering. Recent studies have shown that TGB exhibits this chiral tunneling effect at certain twist angles and gate voltages, providing opportunities for the design of novel electronic devices such as tunnel transistors and tunnel diodes \cite{cao2018unconventional,yankowitz2019tuning,ohta2006controlling,he2013chiral}. Additionally, TGB has been shown to exhibit superconductivity at certain twist angles and doping levels, which could have implications for the development of high-temperature superconductors \cite{cao2018unconventional,chen2021electrically}. Overall, the unique electronic properties of TGB make it a promising material for a wide range of applications in areas such as electronics and quantum computing.\\ 

In this work, we investigate chiral tunneling in twisted graphene bilayer through a double barrier structure, motivated by recent findings developed in \citep{he2013chiral}. To describe our system particles scattered by the double barrier structures, we formulate our model by setting an effective two-band Hamiltonian. We assume that the double barrier potential has a rectangular shape and is infinite in the y-direction. We investigate the transmission probability for this system numerically, considering incident angle, energy, barrier width, and barrier height dependencies. Additionally, we discuss the relationship between the Klein paradox and the transport properties of charge carriers through the double barrier structure. The revelation that the double barrier potential can be utilized to manipulate the transmission has sparked greater interest in TGB and has paved the way for novel methods of producing graphene–based electronic devices.\\

The present paper is organized as follows. In \cref{tsection2}, we formulate our model by setting the effective two-band Hamiltonian system describing particles scattered by double barrier structures.  \Cref{tsection3} is devoted to numerical analysis and discussion of our results from the perspective of possible suitable conditions of the physical parameters. We conclude our results in \cref{tsection4}.\\

\section{THEORETICAL MODEL 
}\label{tsection2}

\begin{figure}[htp]
	\centering
	\subfigure[]{\includegraphics[width=0.23\textwidth]{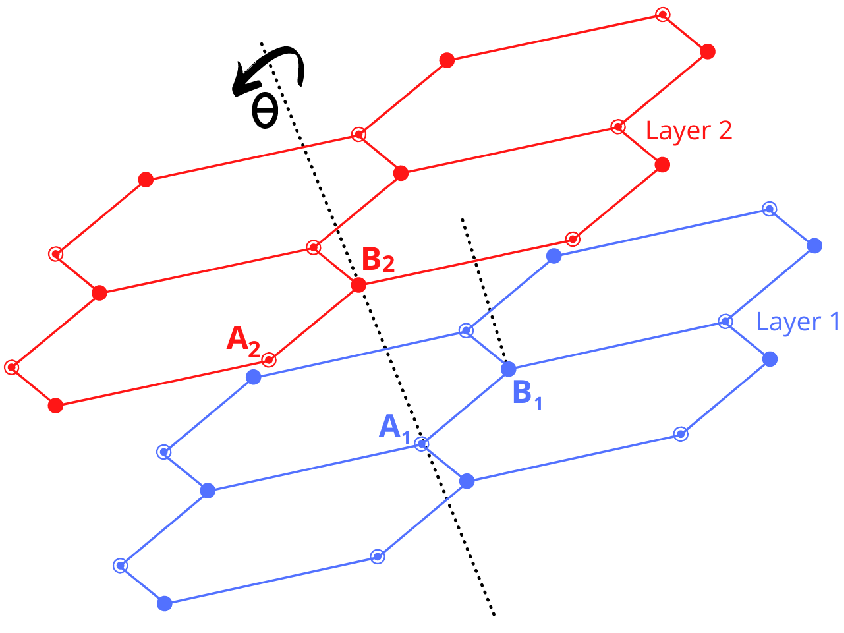}
		\label{fig:2TBGa}} 
	\subfigure[]{\includegraphics[width=0.23\textwidth]{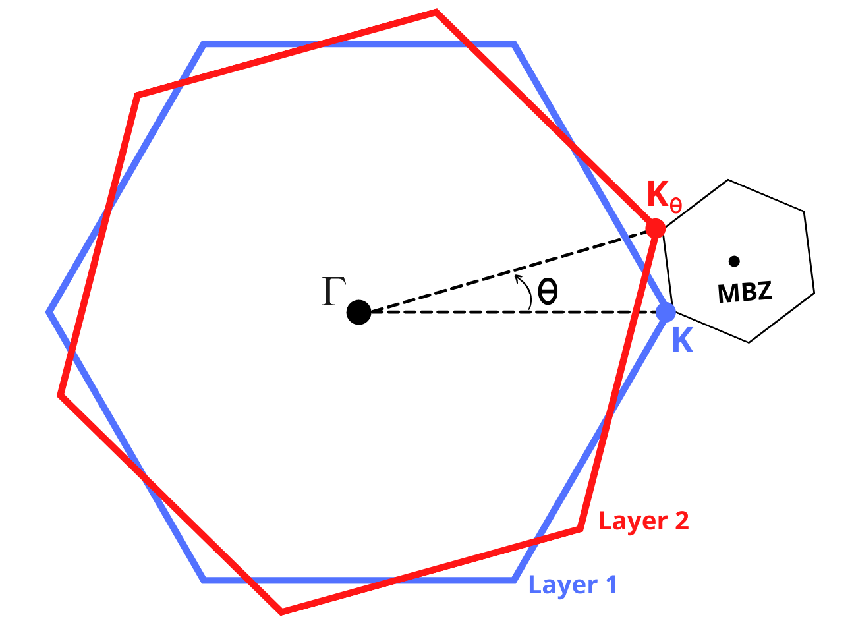}
		\label{fig:2TBGb}}
	\subfigure[]{\includegraphics[width=0.23\textwidth]{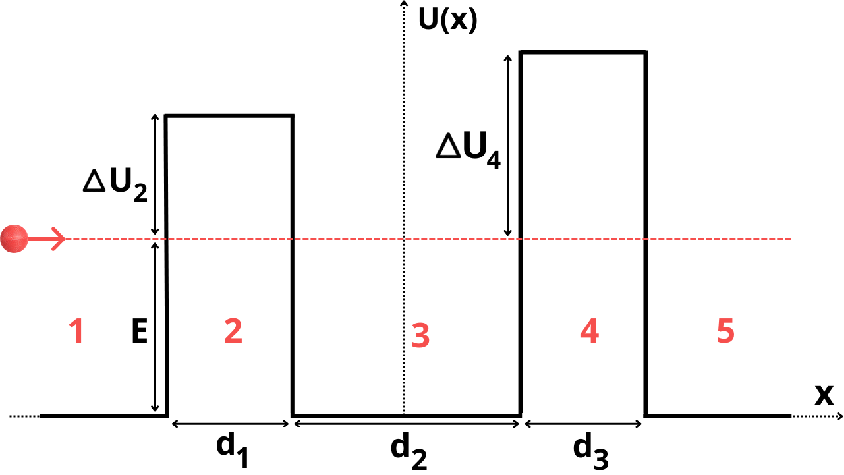}
		\label{fig:1TBGa}}
	\subfigure[]{\includegraphics[width=0.23\textwidth]{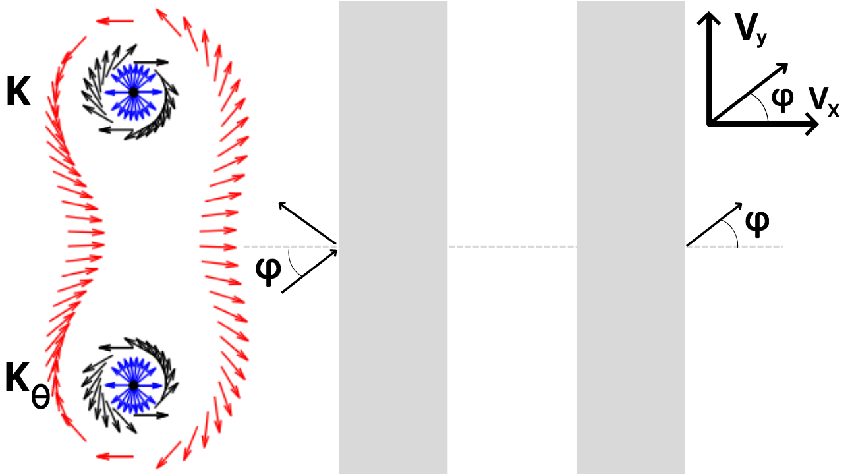}
		\label{fig:1TBGb}}
	{\tiny \caption{\label{fig:1TBG}\textbf{(a)} Schematic representation of TGB can be obtained by rotating the top layer relative to each other by a twist angle $\theta$ around a common $A_{1} B_{2}$ vertical axis. \textbf{(b)} The hexagonal MBZ (black hexagon) of the Moiré superlattice is formed by superposing the BZs of the top (red) and bottom (blue) graphene layers. \textbf{(c)} Schematic diagram for a particle quantum tunneling through a TGB double barrier structure, with the two barriers of width $d_{1 (3)}$, height $U_{2(4)}$ where $U_{2(4)}=E+\Delta U_{2(4)}$, and distance $d_2$ between them.	We assume that the double barrier potential has a rectangular shape and is infinite along the y-direction. \textbf{(d)}  Velocity field of quasiparticles in TGB. The arrows with different colors represent the velocity vectors of quasiparticles with various incident energies $E$. The dark areas are the potential barriers, and $\varphi$ is the incident angle.}}
\end{figure}

The system we study is a twisted graphene bilayer, which is made up of AB-Bernal stacked bilayer graphene with a twist angle $\theta$ between the top and bottom layers, around a common ($A_{1} B_{2}$) vertical axis, as shown in \cref{fig:2TBGa}. Owing to the relative rotation of the Bernal bilayer graphene, a periodic Moiré superlattice is generated and the corresponding Moiré Brillouin zone (MBZ) is depicted schematically in \cref{fig:2TBGb}, along with the Brillouin zones (BZs) of individual graphene layers \cite{li2017splitting}. Moreover, the formation of the Moiré superlattice also directly leads to the expansion of the unit cells of TGB. The twisting not only results in the Moiré pattern, but also leads to a relative shift of the two Dirac cones. The resulting relative shift of the Dirac points on the different layers is defined as $\Delta K$=$2\lvert K\rvert\sin(\theta/2)$, where $\lvert K\rvert$=$4\pi/3a_{0}$ is the momentum-space dependence determined by the graphene lattice constants in real-space $a_{0}=\sqrt{3}a\approx0.25\mathrm{~nm}$. The interaction of electrons between the two graphene sheets and the coupling of their respective Dirac cones in the BZ leads to a redistribution of the electronic band structure, which is similar to that of single-layer graphene with a linear dispersion near the Fermi surface. The electronic band structure of TGB is mainly formed along the anticrossing Dirac cone \cite{cao2018correlated}. The overlap of the electronic band structure between the upper and lower layers of graphene causes the electronic band structure of TGB to exhibit saddle points in the density of state (DOS). These saddle points result in two van Hove singularities (VHSs) for low-energy at $\pm E_{V}$=$\pm 1/2\left(\hbar v_{\mathrm{F}}|\Delta K|-2t_{\perp}\right)$, where $v_{\mathrm{F}}$\;=\;$10^{6}\mathrm{~m/s}$ denotes the Fermi velocity \cite{cao2018correlated,dos2007graphene,dos2012continuum}.\\

To describe the low energy spectrum of TGB electrons, we adopt the continuum nearest-neighbor tight-binding approach \cite{partoens2007normal,neto2009electronic}. The effective Hamiltonian at low-energy excitations \cite{he2013chiral} can be written as :
	\begin{small}
		\begin{equation}
			\label{eqn:h1}
			H^{eff}=-\frac{2 v_{F}^{2}}{15\tilde{t}_{\perp}}\left[\begin{array}{cc}
				0 & \left(k^{\dagger}\right)^{2}-\left(k_{\theta}^{\dagger}\right)^{2} \\
				\left(k\right)^{2}-\left(k_{\theta}\right)^{2}  & 0
			\end{array}\right],       
		\end{equation}
	\end{small}
	where 
	$t_{\perp}$\!\;$\approx$\;0.27eV is the interlayer coupling, which can be approximated  to $\tilde{t}_{\perp}$\!\;$\simeq$ \!0.4$t_{\perp}$ for small twist angles \cite{dos2012continuum}. $k=k_{x}+i k_{y}$ and $k^{\dagger}=k_{x}-i k_{y}$ are the in-plane wave vector and its conjugate, respectively, with $k_{x,y}$=-$i\partial_{x,y}$. For a twist angle $\theta$\!\;$\neq$\;0, we can define the complex wave vector in the MBZ as $k_{\theta}$=$\Delta K/2$=$\left(\Delta K_{x}+i \Delta K_{\mathrm{y}}\right)/2$, and its conjugate as $k_{\theta}^{\dagger}$=$\Delta K^{\dagger}/2$=$\left(\Delta K_{x}-i \Delta K_{y}\right)/2$. 
	For simplicity, we set $\Delta K_{x}=0$ and $\Delta K_{y}=\Delta K$.
	Note that the Hamiltonian's continuum limit is valid for $a\!\ll\!L$, where $L$ is period of the moire patterns in TGB, that is, for small twist angles \cite{dos2007graphene,PhysRevB.84.045436}. The VHSs were detected experimentally in a twisted graphene bilayer with $\theta$\!\;$\leqslant$\!\;$10^{\circ}$ \cite{ohta2012evidence,li2010observation,luican2011single,PhysRevLett.109.126801,PhysRevLett.109.196802,PhysRevB.85.235453,PhysRevLett.108.246103}. As a result, the Hamiltonian (\ref{eqn:h1}) is a good approximation for $\theta$\!\;$\leqslant\!10^{\circ}$ $\left(L\sim1.4\mathrm{~nm}\right.$ for $\left.\theta=10^{\circ}\right)$.\\


In order to study the scattering of Dirac fermions in TGB through a double  barrier structure along the x-direction, we consider the following potential configuration:
	\begin{small}
		\begin{equation}
			U_{j}= 
			\begin{cases}
				E+\Delta U_{2}& \text{if}\; j=2,\\
				E+\Delta U_{4}& \text{if}\; j=4,\\
				0              & \text{elsewhere},
			\end{cases}\label{eqn:p1}
		\end{equation}
	\end{small}
	where $E$ is the incident energy of the electron, $\Delta U_{2,4}$ being the energy difference between the height potential barrier $U_{2,4}$ and the incident energy $E$. The potential double barrier being infinite along the y-direction and can be divided into five different scattering regions $\left(j=1,2,3,4,5\right)$, as shown in \cref{fig:1TBGa}. \\

It is important to note that the group velocity in twisted graphene bilayer, unlike single-layer and Bernal bilayer graphene, is not parallel to the wave vector. It is determined by the fundamental formula $\vec{\nu}_{k}=\hbar^{-1} \left(\nabla_{k} E\right)_{k}$. \Cref{fig:1TBGb} illustrates the velocity field of quasiparticles in twisted graphene bilayer, with arrows of different colors representing the velocity vectors of quasiparticles with various incident energies $E$. Assuming that the two potential barriers have a rectangular shape, the typical width of edge smearing is much smaller than the electron wavelength but much larger than the lattice constant. Therefore, scattering between different valleys cannot mix the two valleys in graphene, and only electron scattering from one valley, specifically from the $K$ and $K{_\theta}$ cones, is considered \cite{katsnelson2006chiral}.\\

According to Eqs. \eqref{eqn:h1} and \eqref{eqn:p1}, we can now write the Hamiltonian describing the five regions as
	\begin{equation}
		\label{eqn:h1new}
		H_{j}=H^{eff}+U_{j}(x) \mathbb{I}_{2}.
	\end{equation}
	Once the corresponding eigenvalue equation is solved, we obtain the energy dispersion relation derived from the Hamiltonian (\ref{eqn:h1new}) as 
	\begin{widetext}
		\begin{equation}
			\label{eqn:h2}
			\epsilon_{j}\left(k_{x}, k_{y}\right)=\pm \frac{2 \nu_{F}^{2}}{15 \tilde{t}_{\perp}} \sqrt{\left(k_{x}^{2}-k_{y}^{2}-\frac{1}{4} \Delta K_{x}^{2}+\frac{1}{4} \Delta K_{y}^{2}\right)^{2}+\left(2 k_{x} k_{y}-\frac{1}{2} \Delta K_{x} \Delta K_{y}\right)^{2}},
		\end{equation}
	\end{widetext}
	where we have set $\epsilon_{j}=E-U_j$.\\

\begin{figure}[htp]
	\centering
	\subfigure[]{\includegraphics[width=0.23\textwidth]{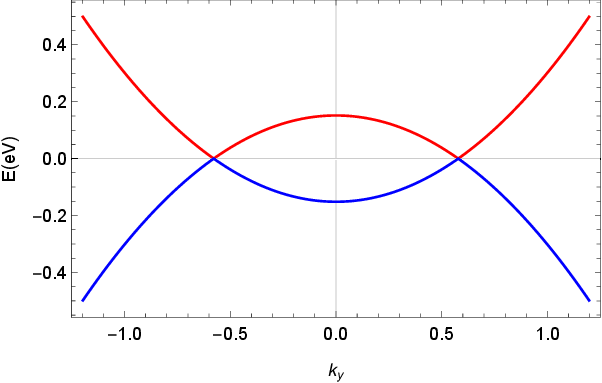}
		\label{fig:2TBGd}}
	\subfigure[]{\includegraphics[width=0.23\textwidth]{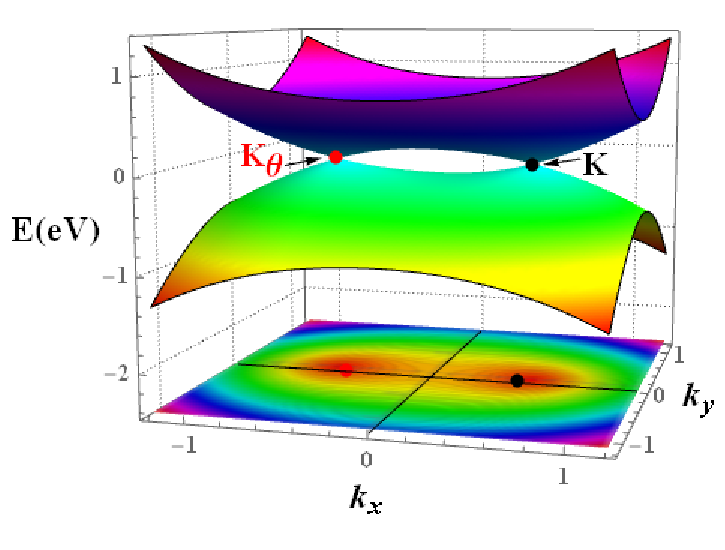}
		\label{fig:2TBGc}}
	{\tiny \caption{\label{fig:2TBG}\textbf{(a)}: The energy spectrum of twisted bilayer graphene near one of the Dirac points for low energy. \textbf{(b)}: The 3D band structure and its projection of the quasiparticles in a twisted graphene bilayer near one of the two valleys for low energy. }}	
\end{figure}

It is clearly seen, as shown in \cref{fig:2TBGc,fig:2TBGd}, that the chiral fermions charged positively and negatively have a symmetrical energy spectrum, and thus we limit our investigation by considering only the nearest neighbor interlayer hopping in the Hamiltonian (\ref{eqn:h1new}). In addition, this symmetry is crucial to study chiral tunneling in most graphene systems, including monolayer graphene, Bernal bilayer graphene, and twisted graphene bilayer \citep{katsnelson2006chiral}. Moreover, the electron-hole symmetry is broken when taking into account large next-nearest-neighbors interlayer hopping, and then it is expected that the chiral tunneling will be destroyed \citep{PhysRevLett.61.2015}. \\

By inserting the full space part of a trial wave function {\footnotesize $\psi_{j}(x, y)$=$\left[\begin{array}{c}\varphi_{j}^{A}(x), \varphi_{j}^{B}(x)\end{array}\right]^{T} e^{i k_{y} y}$} into equation $H_{j} \psi_{j}=E \psi_{j}$ with the Hamiltonian (\ref{eqn:h1new}), we can obtain the eigenstates in the $j$-th region.
	Next, we will calculate the transmission probability of an electron passing through a double barrier from the continuity of the wave functions and their derivatives and look for the effect of incidence angle $\left(\varphi\right)$ and structure parameters $\left(E, U_{2,4}, d_{1,2,3}\right)$ on the tunneling properties through the system. 

\section{RESULTS AND DISCUSSIONS}\label{tsection3}

\begin{figure*}[t]
	\centering
	\includegraphics[scale=0.5]{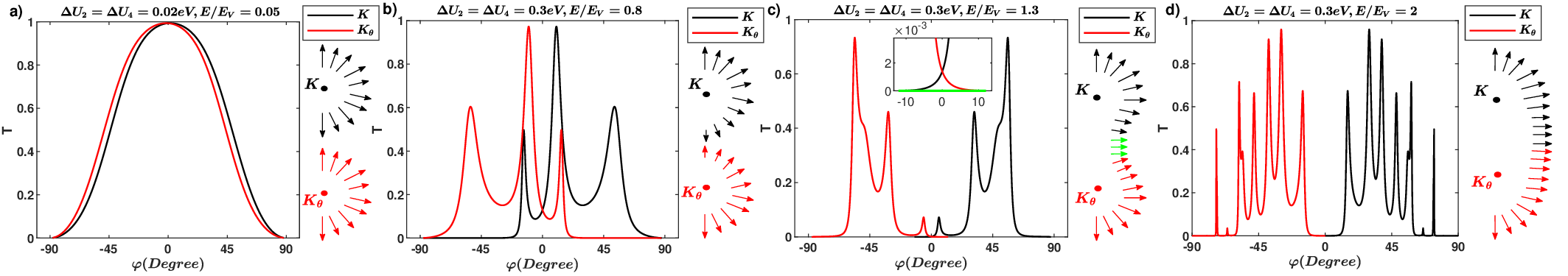} 
	{\tiny 	\caption{\label{fig3:3tbg}The transmission probability in the two cones $K$ (black) and $K_{\theta}$ (red) as a function of the incident angle $\varphi$ for different structure parameters $\left(E, U_{2,4}\right)$. The remaining parameters are the twist angle of the graphene bilayer $\theta=3.89^{\circ}$, $d_{1}=d_{3}=25\ \mathrm{nm}$ and $d_{2}=10\ \mathrm{nm}.$ In \textbf{(c)} The green curve connects the black and red curves, which correspond to the scattered reflection of additional electrons that appear on the symmetry axis directed perpendicular to the line connecting the two cones.}}
\end{figure*}

The dependence of the transmission with respect to the incident angle $T(\varphi)$ for the two cones of the Brillouin zone, i.e., $K$ and $K_{\theta}$, is computed for an electron across a twisted graphene bilayer through double barrier structure and is displayed in \cref{fig3:3tbg} for different incident energies and potential barriers. The velocity fields of quasiparticles in the corresponding energy are also shown in the same figure. The main characteristics of $T(\varphi)$ is that the transmission probability of quasiparticles in the $K$ cone is mirror symmetric with that of the  $K_{\theta}$ cone about $\varphi =0$. As can be seen in \cref{fig3:3tbg}{\color{blue}(a)}, for low energy and a small value of the barrier height, the electrons are transformed into propagating holes inside the barrier,  resulting in perfect transmission. Note, that the Klein tunneling is observed only for normal to near-normal incidence. Similar behavior is also obtained in the case of monolayer graphene \cite{katsnelson2006chiral}.  It is important to note that the normal tunneling is emitted from both $K$ and $K_{\theta}$ cones when the quasiparticles have incident energies less than that of the van Hove singularity, as illustrated in \cref{fig3:3tbg}{\color{blue}(a)} and \cref{fig3:3tbg}{\color{blue}(b)}. Moreover, we observe that the transmission at normal incidence is highly dependent on both the incident energy and the potential barrier heights $\Delta U_2$ and $\Delta U_4$. In fact, by increasing the incident energy and the barrier height, we can clearly see that the transmission changes gradually from perfect transmission, i.e., Klein tunneling, to perfect reflection in the normal direction, which is different from the case of monolayer and AB-stacked bilayer graphene \cite{katsnelson2006chiral}. For the velocity fields of quasiparticles, we note that once the incident energy approaches the van Hove singularity, normal incident electrons from both cones form a semicircle combining them in the reciprocal space. This means that perpendicular incident electrons are emitted from  $K$ cone, $K_{\theta}$ cone and the ridge in the reciprocal space connecting both cones (see \cref{fig3:3tbg}{\color{blue}(c)}). The additional normally incident electrons connecting the $K$ and $K_{\theta}$ is represented by the green color in \cref{fig3:3tbg}{\color{blue}(c)}. The transmission probability, representing these electrons, are always perfectly reflected by the double barrier potential. Moreover, when the incident energy is higher than $2 E_V$, only electrons with wave vector ($k_x$, $0$) are emitted normal to the barrier and the contributions from $K$ and $K_{\theta}$ cones vanish completely. In addition to that, since the transmission probability is always zero for ($k_x$, $0$), the normal tunneling becomes completely forbidden, as we see in \cref{fig3:3tbg}{\color{blue}(d)}. Moreover, it is clearly seen in all figures that the transmission exhibits some sharp resonance peaks at higher angles of incidence. 

\begin{figure*}[!htbp]
	\centering
\includegraphics[scale=0.35]{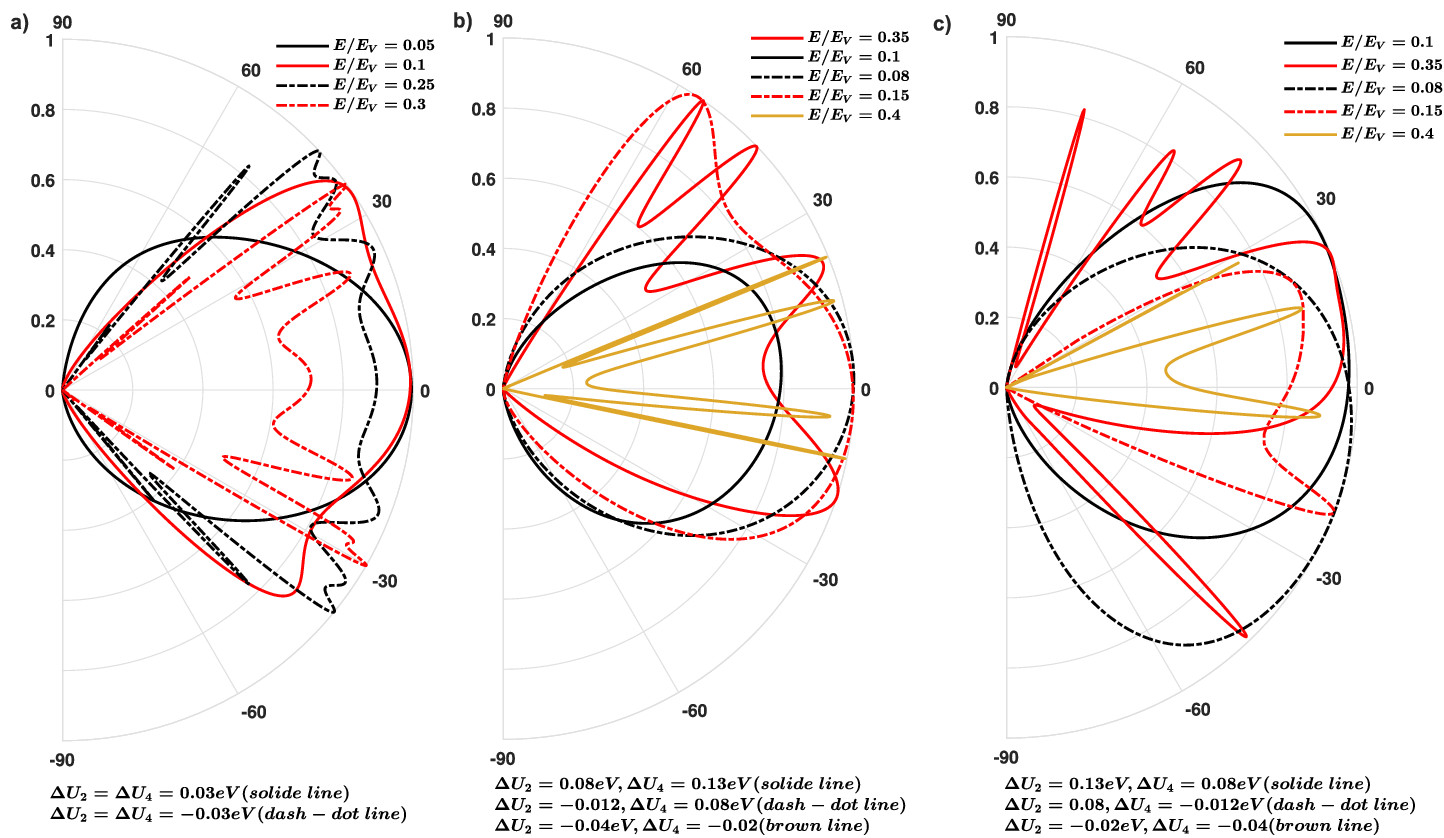} 
{\tiny  \caption{ \label{fig:3TBG} 
Transmission probability $T$ as a function of the incident angle $\varphi$ for symmetric (\textbf{a}) and asymmetric (\textbf{b})-(\textbf{c}) barrier  structures. 
The remaining parameters are the twist angle $\theta=3.89^{\circ}$, $E_{V}=0.15\ \mathrm{eV}$, $d_{1}=d_{3}=100\ nm$ and $d_{2}=10\ nm$. For the case of a symmetric barrier $U_{2}=\!U_{4}$ shown in (\textbf{a}), we plot $T(\varphi)$ for $E\!<\!U_{2,4}$ (solid line) and $E\!>\!U_{2,4}$ (dashed-dotted line). To investigate the effect of barrier structure, we also plot $T(\varphi)$ for asymmetric barrier $U_{2}\neq\!U_{4}$. For $U_{2}<\!U_{4}$ shown in (\textbf{b}), $T(\varphi)$ is plotted for the cases of $E\!<\!U_{2}<\!U_{4}$  (solid line), $U_{2}\!<E<\!U_{4}$ (dashed-dotted line), and $E\!>\!U_{4}>\!U_{2}$ (brown line). For comparison, we consider $U_{2}>\!U_{4}$ shown in (\textbf{c}), we also plot $T(\varphi)$ for the cases of $E\!<\!U_{4}<\!U_{2}$ (solid line), $U_{4}\!<E<\!U_{2}$ (dashed-dotted line), and $E\!>\!U_{2}>\!U_{4}$ (brown line).}}
\end{figure*}
\begin{figure}[!b]
	\centering
	\includegraphics[scale=0.45]{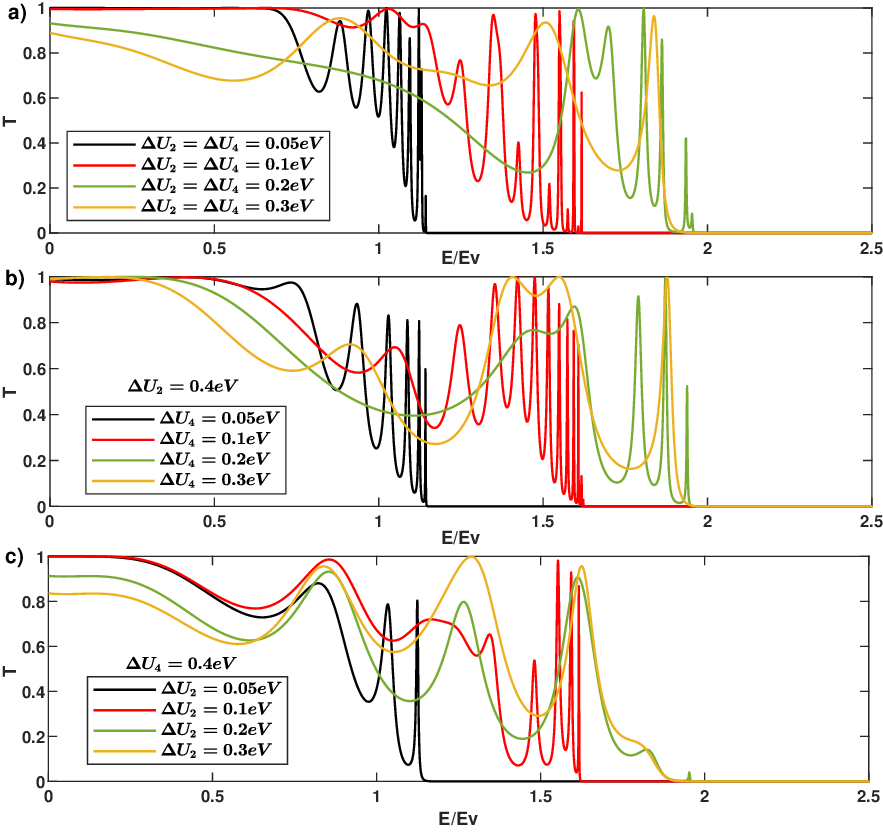} 
{\tiny \caption{\label{fig:3tbg} Transmission probability $T$ for normally incident electrons through a double barrier as a function of incident energy for symmetric (\textbf{a}) and asymmetric (\textbf{b})-(\textbf{c}) barrier structures. The curves with different colors correspond to different values of the potential barrier for all cases. The remaining parameters are the twist angle $\theta=3.89^{\circ}$, $E_{V}=0.15\ \mathrm{eV}$, $d_{1}=d_{3}=100\ nm$ and $d_{2}=10\ nm$.}}
\end{figure}
\Cref{fig:3TBG} presents the transmission dependence of the incident angle $\varphi$ for different cases of $E$ under symmetric and asymmetric barrier conditions ($U_2=U_4$ and $U_2 \neq U_4$). It is evident that the transmission is asymmetric with respect to the x-axis (i.e., normal to the interface). For symmetric barrier and low $E$ with small values of $\Delta U_{2,4}$, we observe perfect tunneling at normal or near normal incidence for $E<U_{2,4}$, as shown in \cref{fig:3TBG}{\color{blue}(a)}. This feature is unique to massless Dirac fermions and is directly related to the Klein paradox. This behavior is similar to that of monolayer graphene for a double barrier structure\cite{BAI20101431, bai2007klein} and in the case of a single barrier for twisted graphene bilayer \cite{he2013chiral}, while the situation is entirely different for the Bernal graphene bilayer structure \cite{bai2007klein}. In the case of monolayer graphene, the propagation of the incident electron wavefunction as hole states inside the barrier results in perfect tunneling. While, in the case of the Bernal graphene bilayer, this propagation transforms into an evanescent wave inside the barrier, and then the chiral electrons are perfectly reflected across a wide barrier \citep{katsnelson2006chiral}.\\
\begin{figure*}[t!]
	\centering
	\includegraphics[scale=0.55]{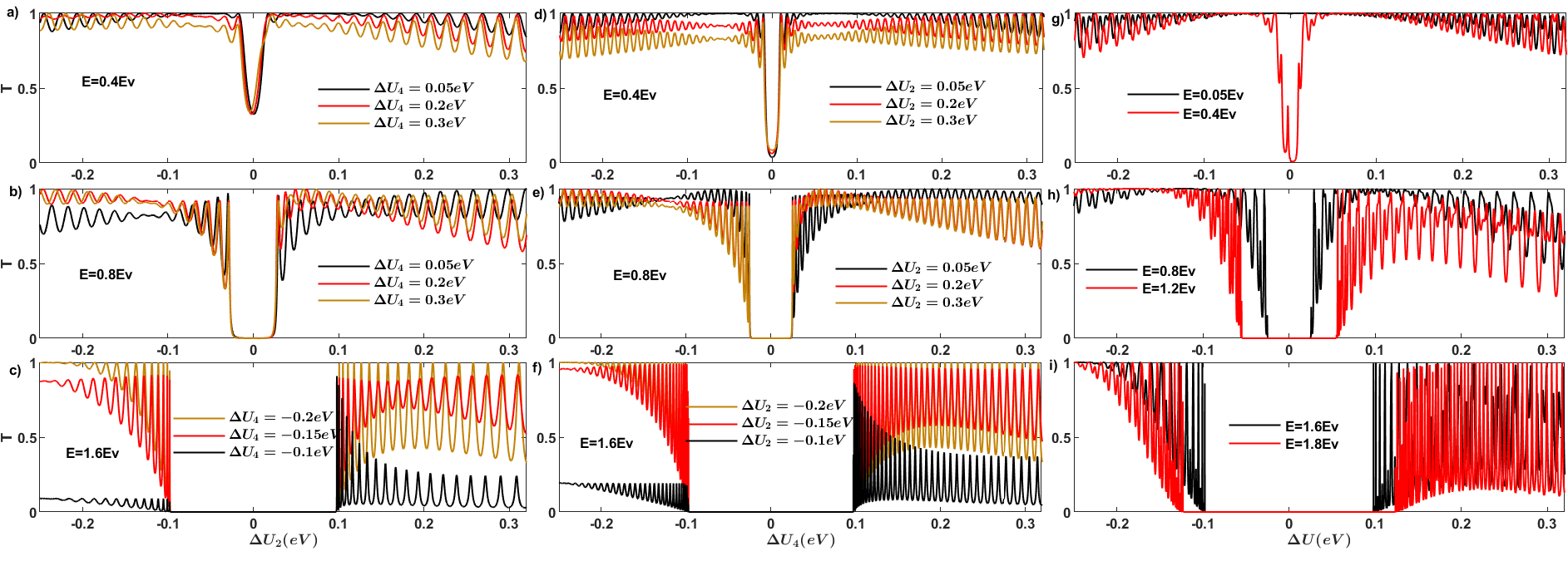} 
{\tiny\caption{\label{fig:5tbg} Transmission probability $T$ for normally incident electrons as a function of $\Delta U_{j}$ ($j=1,\ 2$). The remaining parameters are the twist angle $\theta=3.89^{\circ}$, $E_{V}=0.15\ \mathrm{eV}$, $d_{1}=d_{3}=100\ nm$ and $d_{2}=10\ nm$. (\textbf{a})-(\textbf{c}), transmission probability $T$ as a function of $\Delta U_{2}$ with different values of $\Delta U_{4}$ and $E$. (\textbf{d})-(\textbf{f}), for comparison, we also plot $T$ as a function of $\Delta U_{4}$ with different values of $\Delta U_{2}$ and with the same different values of $E$ used in (\textbf{a})-(\textbf{c}). The curves with different colors in (\textbf{a})-(\textbf{c}) and (\textbf{d})-(\textbf{f}) correspond to different values of $\Delta U_{4}$ and $\Delta U_{2}$, respectively. (\textbf{g})-(\textbf{i}), transmission probability $T$ as a function of $\Delta U\!=\!\Delta U_{2}\!=\!\Delta U_{4}$. Here, The curves with different colors correspond to different values of $E$.}}
\end{figure*}

Furthermore, as the incident energy $E$ increases, some peaks appear, and their number increases with increasing $E$. Moreover, by increasing $E$ the number of those peaks increases, and the amplitude of the transmission decreases systematically before falling rapidly. In addition to that, we can clearly see that the peaks shift towards the higher angle of incidence for symmetric double barrier, i.e. $U_2=U_4$. Furthermore, from \cref{fig:3TBG}{\color{blue}(b)}, we can also find that the structure remains always perfectly transparent for normal or near-normal incidence for $U_{2}<E<U_{4}$, while in the case for $E<U_2<U_4$, $T \neq 1$ for normal incidence. For large value of $E$ ($E>U_{2,4}$) no perfect transmission at normal or near-normal incidence. Additionally, some sharp resonances like peaks occur, and the transmission falls rapidly. In \cref{fig:3TBG}{\color{blue}(c)} we display the transmission dependence of the incident angle $\varphi$ for $U_2>U_4$. From the figure, it may be noted that for low incident energy, the structure is not perfectly transparent for the normal incidence. By increasing $E$ ($E<U_4$), some peaks appear and the amplitude of the transmission decreases systematically before falling rapidly for an incidence angle that is higher than that for a symmetric barrier. When $E> U_2$, the amplitude of the transmission increases, and $T$ falls rapidly as compared to the case for $E<U_4$. It is important to note that the double barrier structure remains in certain cases, regardless of the strength of the potential barrier, perfectly transparent for normal or near-normal incidence.  This is an important characteristic of massless Dirac fermions that is related to the phenomenon of Klein tunneling. The chiral nature of the quasiparticles in graphene causes the tunneling to also be highly dependent on the direction, as previously observed in research studies \cite{ bai2007klein, katsnelson2006chiral, park2008anisotropic}. Remarkably, the anisotropy in the transmission probability through the double barrier structure can be tuned by changing the applied potential strength of the barrier. Furthermore, the height of the double barrier structure provides an additional parameter that allows us to tune the electronic properties of the twisted graphene bilayer. \\

In order to further understand the chiral tunneling in a twisted graphene bilayer through a double potential barrier, we plot the transmission probability for normally incident electrons as a function of incident energy in \cref{fig:3tbg}. For all the cases, the barrier widths are taken to be $d_{1}=d_{3}=100nm$ and the interbarrier separation is $d_{2}=10nm$. The curves with different colors correspond to different values of $\Delta U_{2}$ and $\Delta U_{4}$. It is clearly seen that at low energy, a perfect transmission is obtained for small barrier height (black and red lines) for symmetric and asymmetric barrier. As the energy increases, the transmission remains almost constant and begins to oscillate, with the appearance of some peaks of resonance, before falling rapidly. Moreover, the number and the amplitude of oscillations for symmetric barrier are higher than that for the asymmetric one. It is important to note that by increasing $\Delta U$, the tunneling becomes completely forbidden for a value that shift towards the high energy, but never exceed $2 E_V$. In fact, tunneling becomes completely forbidden for $E\geq 2 E_V$ because the quasiparticles in the twisted graphene bilayer become the pseudospin-1 fermions exactly for $E$ higher than $2 E_V$ \cite{he2013chiral}. Additionally, for large values of $\Delta U$, the number of oscillations decreases as well as the amplitude. Comparing \cref{fig:3tbg}{\color{blue}(a)}, {\color{blue}(b)} and {\color{blue}(c)}, it may be noted that the transmission goes to zero for the same energy if the two barriers, in both symmetric and asymmetric structures, have the same values for the smallest height (represented with the same color).\\


Next, we examine the transmission probability of normally incident electrons in relation to $\Delta U_j$ for symmetric and asymmetric barrier structures with varying incident energy. As shown in \cref{fig:5tbg}, the potential barrier height, represented by $U_j=E+\Delta U_j$, increases from zero as $\Delta U_j$ increases from $-E$. We identify three distinct regions that influence transmission: $ -E< \Delta U<-E_{g}/2$, $|\Delta U|<E_{g}/2$, and $\Delta U > E_{g}/2$, where $E_g$ is the gap between electrons and holes. These regions dictate the transition of the transmission from perfect tunneling to partial reflection, or vice versa.

For low-energy and small $\Delta U$, the upper panels of \cref{fig:5tbg} reveal no transmission gap. However, as the incident energy increases, an energy gap emerges, resulting in complete suppression of transmission ($T=0$) due to the inability to create an electron-hole pair at the barrier interface. The width of the transmission gap also increases with increasing incident energy, as demonstrated in the middle and lower panels. It is worth noting that the width of the transmission gap remains constant for a fixed energy. In addition to that, we can clearly see that the periodicity of the oscillations increases with increasing $\Delta U$ as in the case of single barrier \cite{he2013chiral}. 

For $-E<\Delta U<-E_{g}/2$, the transmission probability oscillates due to the resonance condition and the quasiparticles behave like massive Schrödinger electrons \cite{katsnelson2006chiral}. Meanwhile, for $E>U_{j}$ in  \cref{fig:5tbg}{\color{blue}(a)}, {\color{blue}(b)} and {\color{blue}(c)}, we see that as $U_4$ increases, the transmission amplitude decreases. Moreover, for $\Delta U>E$, as $E$ increases, the periodicity and amplitude of the transmission both increase for both symmetric and asymmetric barriers in the region $\Delta U>E_g/2$. However, for $\Delta U_2<E<\Delta U_4$ in \cref{fig:5tbg}{\color{blue}(f)} and $\Delta U_4<E<\Delta U_2$ in the left lower panel, the transmission oscillates and exhibits maxima, which decrease as $\Delta U_4$ and $\Delta U_2$ increase, respectively.

\begin{figure*}[t]
	\centering
.
	\includegraphics[scale=0.5]{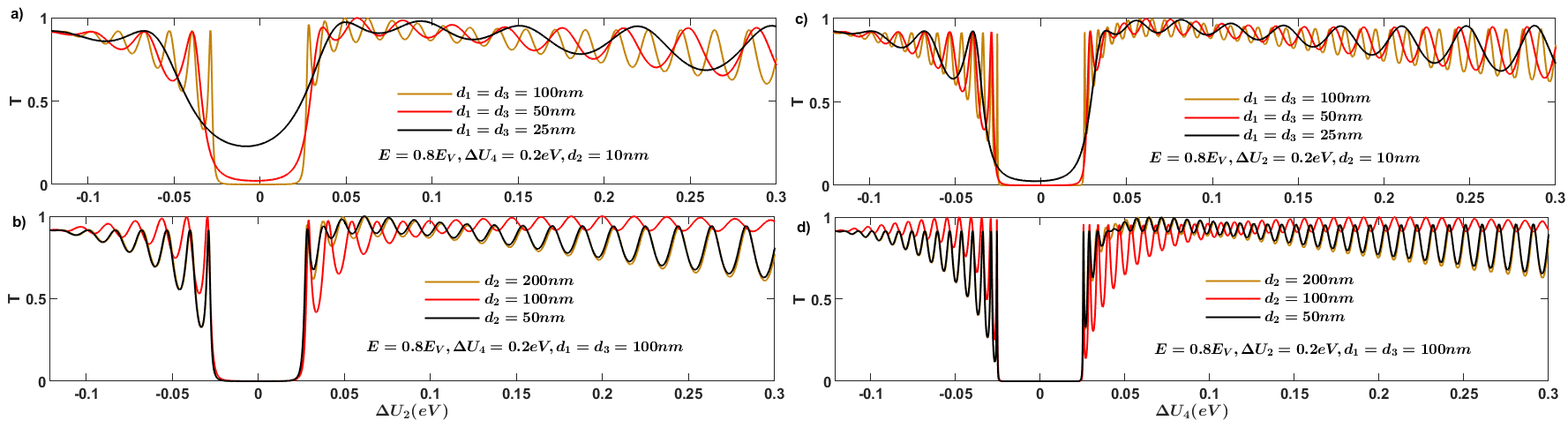} 
{\tiny \caption{\label{fig:8tbg}
		Transmission probability $T$ for normally incident electrons as a function of the barrier heights $\Delta U_{2}$ (\textbf{a})-(\textbf{b}) and $\Delta U_{4}$ (\textbf{c})-(\textbf{d}). The curves with different colors in (\textbf{a}),(\textbf{c}) and (\textbf{b}),(\textbf{d}) correspond to different values of barrier widths $d_{1}\!=\!d_{3}$ and interbarrier separation $d_{2}$, respectively. The remaining parameters are the twist angle $\theta=3.89^{\circ}$ and $E_{V}=0.15\ \mathrm{eV}$.}}
\end{figure*}

In \cref{fig:8tbg}, we show the impact of the barrier width $d$, where $d=d_1=d_3$ as well as the interbarrier separation $d_2$, on the transmission probability as a function of the barrier heights $\Delta U_2$ and $\Delta U_4$.
Specifically, we explore the first barrier height $\Delta U_2$ dependence of transmission probability in \cref{fig:8tbg}{\color{blue}(a)} and {\color{blue}(b)} and the second barrier height $\Delta U_4$ dependence of $T$ in \cref{fig:8tbg}{\color{blue}(c)} and {\color{blue}(d)}. Our results show that, at normal incidence, a transmission gap appears around $E=U$ by increasing $d$. Furthermore, the width of this transmission gap remains constant even when changing $d_2$, as shown in \cref{fig:8tbg}{\color{blue}(b)} and {\color{blue}(d)}. Also, it is clearly seen that the transmission probability oscillates with the barrier height, and the amplitude of the oscillation increases with increasing barrier heights. Additionally, the period of the oscillation increases with decreasing $d$. In fact, the structure is transparent when the electron wave function perfectly matches the wave function for a propagating hole. Furthermore, the energy interval between the nearest states of the propagating hole wave functions inside the barrier is proportional to the height of the barrier. It is worth noting that in the case of monolayer graphene, the double barrier structure is always perfectly transparent for normal incidence, independent of the interbarrier separation or the thickness of the barriers \cite{bai2007klein}. In addition to that, we found that the periodicity of the transmission is more sensitive to the second barrier height, as increasing $\Delta U_4$ results in a larger periodicity of transmission compared to increasing $\Delta U_2$.

\begin{figure}[b!]
	\centering
	\subfigure[]{\includegraphics[width=1.54in,height=1.1in]{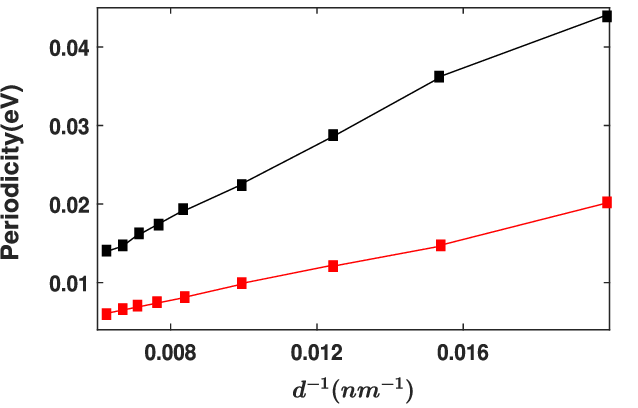}
	\label{fig81:81tbg}}
	\subfigure[]{\includegraphics[width=1.5in,height=1.14in]{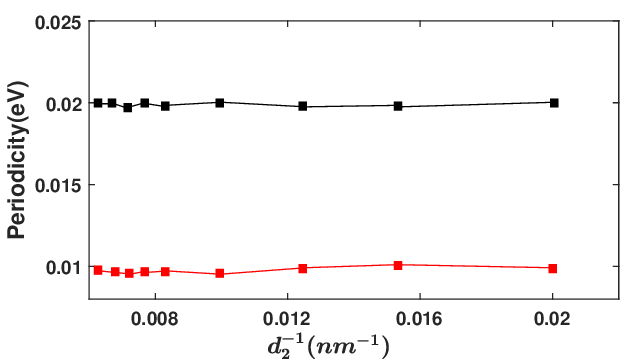}
	\label{fig82:81tbg}}
{\tiny\caption{\label{fig8:81tbg}The average periodicity of the oscillations for $0.3eV <\Delta U_2 <0.4eV$ (\textbf{a}) and $0.3 eV < \Delta U_4 < 0.4 eV$ (\textbf{b}) as a function of $d^{-1}$ and $d_{2}^{-1}$, respectively. Here, $d$ and $d_2$ denote the width of the two barriers $d\!=\!d_1\!=\!d_3$ and the interbarrier separation, respectively. The remaining parameters are the twist angle $\theta=3.89^{\circ}$ and $E_{V}=0.15\ \mathrm{eV}$.}}
\end{figure}

To confirm our previous analysis, we consider the same tunneling problem with different barrier widths $d$, where $d=d_1=d_2$ and the interbarrier separation $d_2$. It is worth noting that as the width of the barrier decreases, the energy gap between the closest states of the hole wave functions propagating within the barrier is expected to increase linearly with the inverse of the barrier width ($d^{-1}$). This implies that the periodicity of oscillations should also rise linearly with $d^{-1}$. Our data, shown in \cref{fig81:81tbg}, confirms that the periodicity of oscillations also increases linearly with $d^{-1}$. Specifically, the periodicity of oscillations with the first and second barrier heights increases linearly with $d^{-1}$. These findings confirm what we observed in \cref{fig:8tbg}. In \cref{fig82:81tbg}, we observe that the periodicity of transmission with the first and second barrier height is not affected by the inverse of the interbarrier separation $d_2$. These results further confirm what we found in \cref{fig:8tbg}. \\

\begin{figure}[t!]
	\includegraphics[scale=0.5]{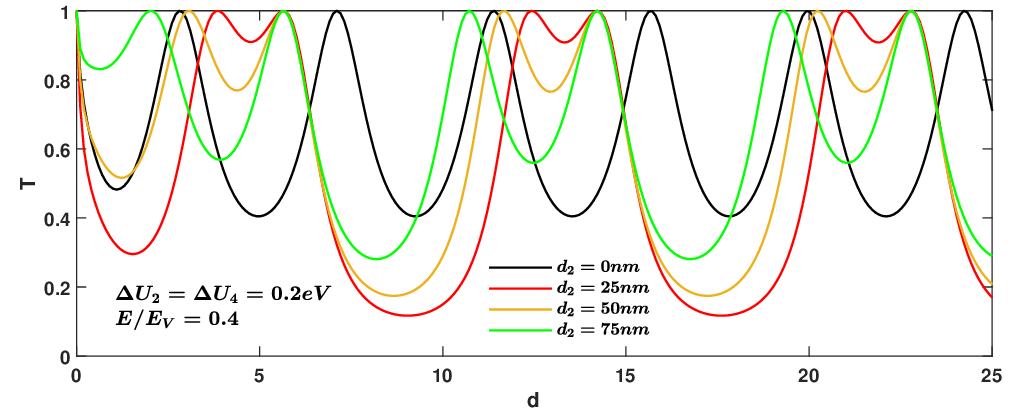}
{\tiny\caption{\label{fig:9tbg}
		Transmission probability of normally incident electrons as a function of barrier width $d\!=\!d_{1}\!=\!d_{3}$. The curves with different colors correspond to different values of interbarrier separation $d_{2}$. The remaining parameters are the twist angle $\theta=3.89^{\circ}$ and $E_{V}=0.15\ \mathrm{eV}$.}}
\end{figure}

In \cref{fig:9tbg}, we plot the transmission probability of twisted graphene bilayer with symmetric double barriers structure as a function of the barrier width $d$, where $d=d_1=d_3$, for different values of the interbarrier separation $d_2$. We can clearly see from this figure that the transmission exhibits oscillatory behavior.  At the same time, we find that amplitude and period of the oscillation depend sensitively on thickness of the barrier and the interbarrier separation. When the interbarrier width is zero, the present structures degenerate into single barrier cases, in which the results (black line) are similar to those found in \cite{he2013chiral}. More remarkably we find that by increasing $d_2$, new oscillations with small amplitude appear, whose amplitude increases by increasing the interbarrier width and the thickness of the barriers.

%

\section{Conclusion} \label{tsection4}

In conclusion, we have investigated the chiral tunneling of the charge carriers in twisted graphene bilayer through double barrier structure. The potential double barrier being infinite along the y-direction and can be divided into five different scattering regions. First, we wrote down the Hamiltonian model that best describes the present system and obtained the relevant energy bands. Then, we looked into the transmission that were examined under various physical parameter circumstances.\\

The transmission probability for both cones, $K$ and $K_{\theta}$, has been computed for different incident energies and potential barriers. According to our numerical analysis, the results showed that the transmission probability of quasiparticles in the $K$ cone is mirror symmetric of that of $K_{\theta}$ cone about $\varphi =0$. Additionally, we found that the normal tunnelling is emitted from both $K$ and $K_{\theta}$ cones when the incident energy is less than that of the van Hove singularity. When the incident energy is higher than $2 E_V$, only electrons with wave vector (kx, 0) are emitted normal to the barrier and the contributions from $K$ and $K_{\theta}$ cones vanish completely. Moreover, our results showed some sharp resonance peaks at higher angles of incidence. In addition to that, we found that in certain cases, regardless of the strength of the potential barrier, the double barrier structure maintains perfect transparency for normal or near-normal incidence. This is an important characteristic of massless Dirac fermions that is related to the phenomenon of Klein tunneling. The chiral nature of the quasiparticles in graphene causes the tunneling to also be highly dependent on the direction. Interestingly, the anisotropy in the transmission probability through the double barrier structure can be tuned by changing the applied potential strength of the barrier. Furthermore, the height of the double barrier structure provides an additional parameter that allows us to tune the electronic properties of the twisted graphene bilayer. These findings provide a better understanding of the electronic properties of twisted graphene bilayer through double barrier structures and can help in designing efficient electronic devices. We have also seen that the amplitude and the periodicity of the transmission depend sensitively on thickness of the barrier and the interbarrier separation. It should be noted that the possibility to control the transmission of charge carriers in twisted graphene bilayer through double barrier structure can help designing efficient graphene-based electronic devices.

\bibliography{TBG}
\bibliographystyle{apsrev4-2}
\end{document}